\documentclass[12pt]{article}
\usepackage{amsmath,amssymb,epsfig}
\makeatletter \@addtoreset{equation}{section} \makeatother
\renewcommand{\theequation}{\thesection.\arabic{equation}}
\newcommand{\ba}{\begin{array}}
\newcommand{\ea}{\end{array}}
\newcommand{\beq}{\begin{equation}}
\newcommand{\eeq}{\end{equation}}
\newcommand{\bea}{\begin{eqnarray}}
\newcommand{\eea}{\end{eqnarray}}
\def\bce{\begin{center}}
\def\ece{\end{center}}

\def\nonu{\nonumber}

\def\be{\beta}

\def\eps6{{\displaystyle \mathop{\epsilon}^{6}}{}}

\def\nab6{{\displaystyle \mathop{\nabla}^{6}}{}}
\def\ft#1#2{{\textstyle{\frac{\scriptstyle #1}{\scriptstyle #2}}}}
\def\fft#1#2{\frac{#1}{#2}}
\def\0{{\sst{(0)}}}
\def\1{{\sst{(1)}}}
\def\2{{\sst{(2)}}}
\def\3{{\sst{(3)}}}
\def\4{{\sst{(4)}}}
\def\5{{\sst{(5)}}}
\def\6{{\sst{(6)}}}
\def\7{{\sst{(7)}}}
\def\8{{\sst{(8)}}}

\def\td{\tilde}
\def\nnn{\nonumber}

\def\ba{\begin{array}}
\def\ea{\end{array}}
\def\beq{\begin{equation}}
\def\eeq{\end{equation}}
\def\be{\begin{equation}}
\def\ee{\end{equation}}

\def\eps{\epsilon}

\def\ba{\begin{array}}
\def\ea{\end{array}}
\def\beq{\begin{equation}}
\def\eeq{\end{equation}}
\def\be{\begin{equation}}
\def\ee{\end{equation}}

\def\eps{\epsilon}

\newcommand{\bean}{\begin{eqnarray*}}
\newcommand{\eean}{\end{eqnarray*}}

\begin{document}
\thispagestyle{empty} \addtocounter{page}{-1}
\begin{flushright}
KIAS-P06003 \\
{\tt hep-th/0603142}\\
\end{flushright}

\vspace*{1.3cm}

\centerline{ \Large \bf From Marginal Deformations to Confinement}
\vspace{.3cm} \centerline{ \Large \bf } \vspace*{1.5cm}
\centerline{{\bf Changhyun Ahn$^{1}$ and {\bf Justin F.
V\'azquez-Poritz}$^{2}$}} \vspace*{1.0cm} \centerline{\it $^{1}$
Department of Physics, Kyungpook National University, Taegu
702-701, Korea} \centerline{\it $^{2}$ Department of Physics,
University of Cincinnati, Cincinnati OH 45221-001, USA }
\vspace*{0.8cm} \centerline{\tt ahn@knu.ac.kr, \qquad
jporitz@physics.uc.edu} \vskip2cm

\centerline{\bf Abstract}
\vspace*{0.5cm}

We consider type IIB supergravity backgrounds which describe
marginal deformations of the Coulomb branch of ${\cal N}=4$ super
Yang-Mills theory with $SO(4)\times SO(2)$ global symmetry. Wilson
loop calculations indicate that certain deformations enhance the
Coulombic attraction between quarks and anti-quarks at the UV
conformal fixed-point. In the IR region, these deformations can
induce a transition to linear confinement.

\baselineskip=18pt
\newpage
\renewcommand{\theequation}
{\arabic{section}\mbox{.}\arabic{equation}}

\section{Introduction}

It has been conjectured that type IIB superstring theory on
$AdS_5\times S^5$ is dual to four-dimensional ${\cal N}=4$ super
Yang Mills theory \cite{malda,GKP,Witten1998}. The supergravity
dual of marginal deformations of this theory related to the
$U(1)\times U(1)$ global symmetry was found in \cite{LM}. This
involved T-dualizing along a $U(1)$ direction of the 5-sphere,
lifting to eleven dimensions and then going back to type IIB
theory along $U(1)$ directions which are shifted by an $SL(3,R)$
transformation \cite{LM}. So long as the direction corresponding
to the $U(1)$ R-symmetry of the theory is not involved in this
procedure, the deformed theory preserves ${\cal N}=1$
supersymmetry. Matching this deformation of $AdS_5\times S^5$ to
an exactly marginal operator in the field theory provides a
holographic test of the methods in \cite{LS}. In this manner,
marginal deformations have been found for various superconformal
field theories, including those associated with the conifold as
well as the $Y^{p,q}$ spaces \cite{LM} and the $L^{p,q,r}$ spaces
\cite{AV}\footnote{The Sasaki-Einstein spaces $Y^{p,q}$ and
$L^{p,q,r}$ were found in \cite{Ypq1,Ypq2} and \cite{Lpqr1,Lpqr2},
respectively.}. Also, the eleven-dimensional supergravity duals of
marginal deformations of various three-dimensional conformal field
theories were found in \cite{AV,GLMW}.

This method can also be used to find the gravity duals of field
theories undergoing renormalization group (RG) flows
\cite{LM,nunez,ahn}. In particular, the gravity dual description
of deformations of the Coulomb branch of ${\cal N}=4$ Yang Mills
theory with $SO(2)^3$ global symmetry was considered in
\cite{ahn}. This part of the moduli space corresponds to a
continuous distribution of D3-branes on an ellipsoidal shell
\cite{larsen}. The introduction of scalar expectation values does
not change the superpotential of the theory. Furthermore, the
superpotential along the deformed Coulomb branch flow is the same
as that for the marginal deformations at the UV conformal fixed
point \cite{LM}, and is given by
%%%%%%%%%%%%%%%
\bea W = \mbox{Tr} \left( \Phi_1 \Phi_2 \Phi_3 -q \Phi_2 \Phi_1
\Phi_3 \right), \qquad q^n =1\,,\eea
%%%%%%%%%%%%%%%
which is a $q$-deformation of the superpotential preserving ${\cal
N}=1$ supersymmetry \cite{LS}. $\Phi_i$ are three adjoint chiral
superfields, $q=e^{2\pi i (\gamma-\tau_s \sigma)}$ and $\tau_s$ is
related to the gauge coupling and theta parameter of the field
theory.

The setup can be pictured as in Figure 1. The ${\cal N}=4$ super
Yang Mills theory undergoes the RG flow on the left when certain
$SO(6)$ scalar fields acquire non-vanishing expectation values. We
will focus on a part of the moduli space for which Wilson loop
calculations \cite{rey,malda2} indicate that there is a Coulombic
force between quarks and anti-quarks \cite{sfetsos1}. After
applying $\sigma$ deformations to the UV conformal fixed point,
the theory undergoes a different RG flow, as shown on the right of
Figure 1\footnote{This picture is a bit over-simplified, since
there is actually a countably infinite family of deformations that
can be applied at all points along the original RG flow.}. How do
the marginal deformations of the UV conformal fixed point effect
the physics of the IR region? From the Wilson loops, we find that
these $\sigma$ deformations can actually cause the theory to
undergo a phase transition such that the quark anti-quark pairs
exhibit linear confinement\footnote{Wilson loops for string
configurations which are not purely radially oriented have been
studied for the $\gamma$ deformed backgrounds \cite{sfetsos}.}.
Similar phenomena were discussed on the gauge theory side in
\cite{dorey1,dorey2,dorey3}. We should note that we only consider
external quarks.

%%%%%%%%%
\begin{figure}[ht]
   \epsfxsize=4.0in \centerline{\epsffile{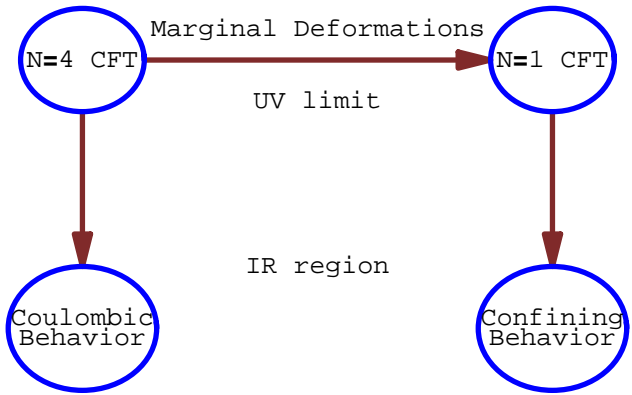}}
   \caption[FIG. \arabic{figure}.]{Marginally deformations can cause
   conformal field theories to undergo different Renormalization Group flows.
   The IR region of these theories can be drastically changed.
   We will consider examples in which $\sigma$ deformations of the
   UV conformal fixed point induces a transition in the
   IR region from Coulombic behavior to linear confinement.}
   \label{coulomb3}
\end{figure}
%%%%%%%%%

This paper is organized as follows. In section 2, we show that the
marginal deformations of ${\cal N}=4$ super Yang Mills theory
enhance the Coulombic attraction between quarks and anti-quarks.
In section 3, show that deformations along certain parts of the
moduli space of the Coulomb branch cause a phase transition to a
linearly confining theory. We discuss various issues and further
directions in section 4.

\section{Marginal deformations of AdS$_5\times S^5$}

For the AdS$_5\times S^5$ background of type IIB theory, Wilson
loop calculations indicate that the energy of a string stretched
along the radial direction is $E=c/L$, where $L$ is the distance
between the string endpoints and the strength parameter $c$ is
proportional to $\sqrt{g_{YM}^2N}$ \cite{malda2,rey}. This is to
be expected from conformal invariance of the dual four-dimensional
gauge theory \cite{malda}\footnote{This carries over when $S^5$ is
replaced by $T^{1,1}$ or one of the countably infinite $Y^{p,q}$
\cite{Ypq1,Ypq2} or $L^{p,q,r}$ spaces \cite{Lpqr1,Lpqr2}.}. The
Lunin-Maldacena deformation of this background has the
string-frame metric \cite{LM}
%%%%
\be ds_{str}^2 = \alpha^{\prime} \sqrt{H}
(\fft{r^2}{R^2}dx_{\mu}^2+R^2 \fft{dr^2}{r^2}+R^2 ds_{{\td
S}^5})\,, \ee
%%%%
where the characteristic length scale of AdS$_5$ is given by
$R^4=4\pi g_{YM}^2 N$ in string units. ${\td S}^5$ is the
deformation of $S^5$ which depends on the modulus of
$\beta=\gamma-\tau_s \sigma$, where $\gamma$ and $\sigma$ are real
deformation parameters and $\tau_s$ is a complex structure
parameter related to the gauge coupling and theta parameter of the
dual gauge theory \cite{LM}. The conformal factor $H$ is given by
%%%%
\be H=1+4\hat\sigma^2\, s_{\alpha}^2 (c_{\alpha}^2+s_{\alpha}^2
s_{\theta}^2 c_{\theta}^2)\,, \ee
%%%%
where $\alpha$ and $\theta$ are internal coordinates of ${\td
S}^5$ and $\hat\gamma\equiv \gamma R^2/2$, $\hat\sigma\equiv
\sigma R^2/2$. The classical supergravity description is valid as
long as the curvature is small relative to the string scale, the
two-torus corresponding to the $U(1)^2$ global symmetry is larger
than the string scale, and the metric does not degenerate at
arbitrary points. These conditions can be met for \cite{LM}
%%%%
\be R>>1\,,\qquad \hat\gamma<<R\,,\qquad \hat\sigma<<R\,.
\label{sugraconditions} \ee
%%%%

From the AdS/CFT correspondence, the gluonic field created by a
static quark anti-quark pair can be described by a string
configuration whose endpoints are located on the boundary of the
AdS spacetime \cite{rey,malda2}. The Wilson loop can be computed
by minimizing the Nambu-Goto action for a fundamental string on
the supergravity background. The action for the Euclideanized
string worldsheet is given by
%%%%
\be S=\fft{1}{2\pi\alpha^{\prime}} \int d\tau\, d\sigma\,
\sqrt{{\rm det} G_{MN}\partial_{\alpha} X^M \partial_{\beta}
X^N}\,. \ee
%%%%
For a static configuration, we can take $\tau=t$ and $\sigma=x_1$.
Also, because our configuration is static, the boundary action due
to the antisymmetric tensor field vanishes does not contribute to
the string action.

In order for a purely radial string to solve the equations of
motion, it is required that
$\partial_{\alpha}H=\partial_{\theta}H=0$. For example,
$H=1+\hat\sigma^2$ for $\theta=0$ and $\alpha=\pi/4$, as well as
$\theta=\pi/4$ and $\alpha=\pi/2$. Also, $H=1+\ft43 \hat\sigma^2$
for $\theta=\pi/4$ and $c_{\alpha}=1/\sqrt{3}$. Then
%%%%
\be S=\fft{T\sqrt{H}}{2\pi} \int dx_1 \sqrt{(\partial_{x_1}r)^2
+r^4/R^4}\,. \ee
%%%%
Since $H$ is a constant, the Wilson loop calculation proceeds as
in \cite{malda2} with the result that the energy of the string
configuration is
%%%%
\be E=-\fft{4\pi^2\sqrt{2g_{YM}^2 NH}}{\Gamma(1/4)^4\, L}\,, \ee
%%%%
where $L$ is the distance between the string endpoints. We will
show the intermediate steps of such computations in the next
section, which includes the above result in the special case of
vanishing $\ell_i$. Since this deformation did not break conformal
invariance, the energy goes as $1/L$, since the only scale present
is $L$. However, the charges are effectively enhanced by the
factor $\sqrt{H}$. Notice that the $\gamma$ deformations have no
effect in this regard, so long as the string is oriented purely in
the radial direction.

\section{Deformations of Coulomb branch flows}

We would now like to investigate the effects of these deformations
on the quark anti-quark potential for nonconformal field theories,
in particular with regards to confinement. The natural place to
begin such endeavors is to consider deformations of theories which
already exhibit confinement. It turns out that for the
Klebanov-Strassler \cite{KS} and Maldacena-N\'u$\td{\rm n}$ez
\cite{MN} backgrounds, both of which exhibit confining IR phases,
the quark anti-quark potential remains invariant under $\gamma$
deformations. From the last section, we might wonder if $\sigma$
deformations might enhance the scale of confinement. However,
these backgrounds are not well-defined under $\sigma$ deformations
\cite{LM,nunez}. We do not consider these types of deformations of
the Polchinski-Strassler solution \cite{PS} because they would
break all of the R-symmetry and therefore not be supersymmetric.
In addition, since it is a perturbative solution, we are not
guaranteed that these deformations would maintain the regularity
of the solution.

The quark anti-quark potential has been studied for the $\gamma$
deformed Coulomb branch of ${\cal N}=4$ super Yang Mills theory
\cite{sfetsos}, where $\gamma$ deformations were found to effect
the behavior only when the two $U(1)$ factors in the global gauge
group corresponding to the quark and anti-quark are given
different expectation values. We would like to consider $\sigma$
deformations of this theory. The supergravity dual of the deformed
Coulomb branch flow can be written (with the metric expressed in
the string frame) as \cite{ahn}
%%%%
\bea ds^2 &=& \alpha^{\prime}\sqrt{Hf}R^2\Big[ \fft{G}{g}
s_{\alpha}^2 (D\varphi_1+gL_3
c_{\theta}^2\,D\varphi_2)^2+\fft{G}{4h} D\varphi_2^2
+\fft{r^2}{fR^4}
dx_{\mu}^2 +\fft{dr^2}{f r^2 L_1 L_2 L_3}\nnn\\
&+& k^{-1}\, [d\alpha+\fft{k}{4}(L_2-L_3)s_{2\alpha}
s_{2\theta}\,d\theta]^2 + kf^{-1} s_{\alpha}^2\,d\theta^2
+\fft{9}{4} ghL_1 L_2 L_3 s_{2\alpha}^2 s_{2\theta}^2\,d\psi^2
\Big] \,,\nnn\\
F_{(5)} &=& dC_{(4)}+\ast\, dC_{(4)}\,,\qquad C_{(4)} = \Big(
f^{-1} \fft{r^4}{R^4} -1\Big) \,d^4x\,,\nnn\\
B & = & \gamma\, \fft{R^4 f\,s_{\alpha}^2}{4gh}\, G\, D \varphi_1
\wedge D \varphi_2 -\sigma\, {\cal A}_{(2)}\,, \qquad C^{(2)}  =
-\sigma\, \fft{R^4 f\,s_{\alpha}^2}{4gh}\, G\, D \varphi_1 \wedge
D \varphi_2 -\gamma\, {\cal A}_{(2)}\,,
\nonu \\
e^{2\phi} & = & H^2 G\,,\qquad \chi  =  \gamma \sigma\, \fft{R^4
f\,s_{\alpha}^2}{4ghH}\,, \label{background} \eea
%%%%
where
%%%%
\bea d{\cal A}_{(2)} &=& \fft34 R^4 \fft{f^2}{r} s_{2\alpha}
s_{2\theta} \,d\psi\wedge \Big( \fft{L_1L_2L_3}{r^2}
\partial_r (f^{-1} r^4) s_{\alpha}^2\,
d\alpha\wedge d\theta + k f^{-1}
\partial_{\alpha} (f^{-1}) s_{\alpha}^2\,
dr\wedge d\theta\nnn\\ &+& [\fft{L_3-L_2}{4} s_{2\alpha}
s_{2\theta} \partial_{\alpha}(f^{-1})+\fft{1}{k}
\partial_{\theta}(f^{-1})]\, dr\wedge [d\alpha+\fft{k}{4}(L_2-L_3)
s_{2\alpha} s_{2\theta}\,d\theta] \Big)\,, \eea
%%%%
and $D\varphi_i= d\varphi_i+A_{\psi}^i d\psi$ with the connection
one-forms
%%%%%%%%%%%
\be A_{\psi}^1 = [L_3 c_{\theta}^2(1-A_{\psi}^2)-L_2
s_{\theta}^2]g\,,\qquad A_{\psi}^2 = (2gL_2 L_3 s_{\alpha}^2
s_{2\theta}^2-4L_1 c_{\alpha}^2)h\,. \ee
%%%%
The various functions are given by
%%%%
\bea f^{-1} &=& (L_1^{-1} c_{\alpha}^2+L_2^{-1} s_{\alpha}^2
s_{\theta}^2+L_3^{-1} s_{\alpha}^2 c_{\theta}^2)L_1 L_2 L_3
\,,\qquad g^{-1} = L_2 s_{\theta}^2+L_3 c_{\theta}^2 \,,\nnn\\
h^{-1} &=& 4L_1 c_{\alpha}^2+gL_2 L_3 s_{\alpha}^2
s_{2\theta}^2\,,\qquad k^{-1}=L_1 s_{\alpha}^2+g^{-1}
c_{\alpha}^2\,, \eea
%%%%
and
%%%%
\be L_i = 1+\fft{\ell_i^2}{r^2}\,. \ee
%%%%
The deformation functions are
%%%%
\be G^{-1}=1+(\hat\gamma^2+\hat\sigma^2)\fft{f}{gh}
s_{\alpha}^2\,,\qquad H=1+\hat\sigma^2 \fft{f}{gh} s_{\alpha}^2\,.
\label{G} \ee
%%%%
The deformations turn on the complex 3-form field strength, which
supports D5 and NS5-branes wrapped on a two-torus. As discussed in
\cite{LM}, in order for the 5-brane charges to be quantized,
$\gamma$ and $\sigma$ must take on rational values. Also, since
the dilaton is no longer constant, the corresponding super
Yang-Mills coupling constant runs along the flow.

The deformed geometry is regular \cite{ahn}, with the exception of
the singular shell of D3-branes which may represent a phase
transition at the associated scale of the RG flow \cite{larsen}.
The conditions for the classical supergravity description to be
valid include those for the marginal deformations of $AdS_5\times
S^5$, and are given by (\ref{sugraconditions}). Even so, the
curvature blows up as one approaches the singular shell of
D3-branes which, from the point of view of the dual gauge theory,
may correspond to a phase transition \cite{larsen}. This is not a
problem for us as long as the curvature is small at the minimal
radius $r_0$ which is probed by string configurations. This is
guaranteed if $r_0>> \ell_i/R^2$. All of these conditions are met
for large enough $R$.

We will now consider inserting a probe D3-brane at large distance
and consider the behavior of string configurations which end on
this brane. For the purposes of Wilson loop calculations, only the
metric is important. Since we will only consider static string
configurations, antisymmetric tensor fields do not contribution to
the string action. The symmetry in the $x_i$ directions allows us
to choose coordinates so that the trajectory runs in the
$x_1\equiv x$ direction with $x_2=x_3=0$. Also, since the metric
is independent of $\varphi_i$ and $\psi$, taking them to be
constant is automatically consistent with the corresponding
equations of motion. With these simplifications, the string action
is
%%%%
\be S=\fft{T}{2\pi} \int dx \sqrt{H}
\sqrt{\fft{r^4}{fR^4}+\fft{r^{\prime 2}}{fL_1L_2L_3} +\fft{r^2}{k}
\alpha^{\prime 2} +\fft{r^2 k}{f} s_{\alpha}^2 \theta^{\prime
2}+\fft{r^2}{2} (L_2-L_3)s_{2\alpha} s_{2\theta}
\alpha^{\prime}\theta^{\prime}}\,, \ee
%%%%
where $^{\prime}$ denotes a derivative with respect to $x$ and $T$
denotes the time interval. A simple solution to the equations of
motion is given by $\alpha=\pi/2$, $\theta=\pi/4$ and
$\ell_2=\ell_3$. This restricts us to the case with $SO(4)\times
SO(2)$ symmetry. The trajectories are parallel to an axis that
passes through the center of the D3-brane distribution. Notice
that, for a spherical distribution of branes, all $\ell_i$ are
equal and $H$ reduces the constant values of the previous section.
Since the action does not explicitly depend on $x_1$, the solution
satisfies
%%%%
\be \sqrt{\fft{H}{f}}
\fft{r^4}{\sqrt{\fft{r^4}{R^4}+\fft{r^{\prime
2}}{L_1L_2^2}}}=\sqrt{\fft{H_0}{f_0}} r_0^2 R^2\,, \ee
%%%%
where $H_0\equiv H(r_0)$, $f_0\equiv f(r_0)$ and $r_0$ is the
minimum value of $r$ that the string reaches. The solution is
given by
%%%%
\be x= R^2 \int_{r_0}^r \fft{dr}{r^2 \sqrt{L_1}L_2
\sqrt{\fft{f_0Hr^4}{fH_0r_0^4}-1}}\,, \label{x1} \ee
%%%%
where $r_0$ is determined by $L=2 x(r\rightarrow \infty)$.

Figure 2 shows string configurations for $\ell_1=0$,
$\ell_2=\ell_3=10$. The plots for $\hat\sigma=1$ (black), $2$
(red) and $3$ (blue) serve to demonstrate that, at a given $L$,
the $\sigma$ deformations increase the energy scale ($r$) probed
by a string. One condition in order for the $\sigma$ deformations
to induce a phase transition to a confining phase is that the
energy scale must be enhanced all the more for large $L$.
%%%%%%%%%
\begin{figure}[ht]
   \epsfxsize=4.0in \centerline{\epsffile{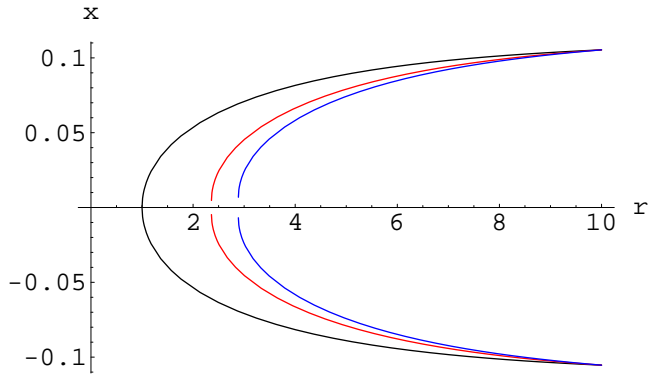}}
   \caption[FIG. \arabic{figure}.]{String configuration for $\ell_1=0$,
   $\ell_2=\ell_3=10$ and $\hat\sigma=1$ (black), $2$ (red) and $3$ (blue).
   For a given $L$, the $\sigma$ deformations increase the energy scale probed by a string.}
   \label{coulomb4}
\end{figure}
%%%%%%%%%

The regularized energy of the string configuration is
%%%%
\be E=\fft{1}{\pi} \int_{r_0}^{\infty} dr\Big[
\sqrt{\fft{f_0}{H_0}}\fft{Hr^2}{fr_0^2 \sqrt{L_1} L_2
\sqrt{\fft{f_0Hr^4}{fH_0r_0^4}-1}}- \sqrt{1+\hat\sigma^2}\Big]
-\fft{1}{\pi} \int_0^{r_0} dr \sqrt{1+\hat\sigma^2}\,, \label{E}
\ee
%%%%
where the infinite contribution from the W-bosons has been
subtracted. $L$ and $E$ can be written in terms of complete
elliptic integrals for two separate cases \cite{gr}.

The first case is given by
%%%%
\bea L &=& 4R^2 (1+\hat\sigma^2)\beta_1
\sqrt{(r_0^2+\ell_2^2)\alpha_1}\, [\Pi
(\alpha_1,\sqrt{\alpha_2})-K(\sqrt{\alpha_2})]\,,\nnn\\
E &=& \frac{\sqrt{\beta_2}}{2\pi} \left[ \frac{\alpha_1}{\beta_2}
K(\sqrt{\alpha_2}) -\frac{(1+\hat{\sigma}^2)
(r_0^2+\ell_1^2)}{1-\alpha_2} E\left(\sqrt{\alpha_2} \right)
\right]\,. \label{case1} \eea
%%%%
where\footnote{We denote $K, E$, and $\Pi$ for the complete
elliptic integrals of the first, second and third kind,
respectively.}
%%%%
\be \alpha_i \equiv [r_0^2+\ell_i^2+\hat\sigma^2
(r_0^2+2\ell_2^2-\ell_{3-i}^2)]\beta_2\,,\qquad \beta_i^{-1}\equiv
2r_0^2+\ell_1^2+\ell_i^2+2\hat\sigma^2 (r_0^2+\ell_2^2)\,. \ee
%%%%
This case applies when either $r_0^2\ge \ell_1^2-2\ell_2^2$ for
any $\hat\sigma$, or else $r_0^2<\ell_1^2-2\ell_2^2$ and
$\hat\sigma^2<\ft{r_0^2+\ell_2^2}{\ell_1^2-2\ell_2^2-r_0^2}$. For
equal $\ell_i$, $L$ becomes independent of $\hat\sigma$ and
reduces to that of the conformal case considered in \cite{malda2}.
For vanishing $\hat\sigma$ this reduces to the undeformed Coulomb
branch, particular cases of which have been considered in
\cite{sfetsos1}.

Figures 3 and 4 are parametric plots of the distance between the
quark and the anti-quark $L$ versus the quark anti-quark potential
$E$ using (\ref{case1}) for case 1. The effect of the $\sigma$
deformations depends on which part of the Coulomb branch we are
probing, as well as the trajectories of the probe strings. For
trajectories which are orthogonal to a uniform distribution of
D3-branes on a two-dimensional disk of radius $\ell_1$ ($\ell_1\ne
0$, $\ell_2=\ell_3=0$), we find that the $\sigma$ deformations
enhance the attractive Coulombic force between the quark and
anti-quark. This force vanishes at asymptotically large
distance\footnote{Our results indicate that there is complete
screening for large enough $L$, as was found for these
trajectories in the undeformed backgrounds \cite{sfetsos1}.
However, this occurs only for strings which touch the D3-brane
distribution and is simply a remnant of the continuous
approximation of the distribution.}. This is shown in Figure 3 for
$\ell_1=1$ and $\hat\sigma=0$, $10$ and $20$. We have added
$\sigma$ dependent constants to the energy (since energy is only
defined up to an additive constant) in order to render it zero
when the quark anti-quark distance gets very large. We have set
$R$ to unity for convenience, since it simply has the effect of
rescaling $\hat\sigma$ and $L$.

%%%%%%%%%
\begin{figure}[ht]
   \epsfxsize=4.0in \centerline{\epsffile{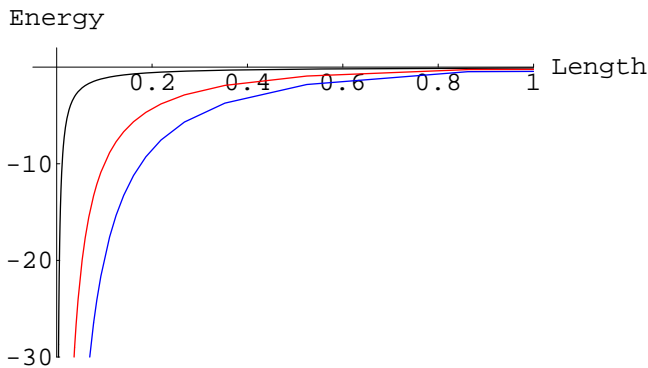}}
   \caption[FIG. \arabic{figure}.]{Quark anti-quark potential using
   (\ref{case1}) for $\ell_1=1$, $\ell_2=\ell_3=0$ and $\hat\sigma=0$
   (black), $10$ (red) and $20$ (blue).
   In this part of the Coulomb branch, the $\sigma$ deformations simply
   enhance the Coulombic force.}
   \label{coulomb1}
\end{figure}
%%%%%%%%%

%%%%%%%%%
\begin{figure}[ht]
   \epsfxsize=4.0in \centerline{\epsffile{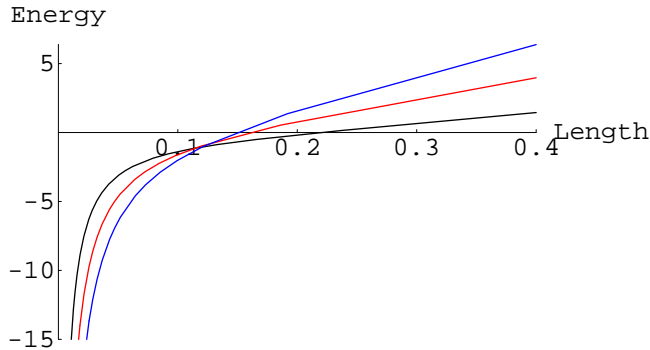}}
   \caption[FIG. \arabic{figure}.]{Quark anti-quark potential using
   (\ref{case1}) for $\ell_1=0$, $\ell_2=\ell_3=10$ and $\hat\sigma=1$
   (black), $2$ (red) and 3 (blue). $R$ has been set to unity. For
   small distance, the force is Coulombic. As the distance
   increases in this part of the Coulomb branch, the $\sigma$ deformations
   induce a transition to a linearly confining phase.
   The scale of confinement increases with $\sigma$.}
   \label{coulomb2}
\end{figure}
%%%%%%%%%

For a uniform distribution of D3-branes on a three-dimensional
spherical shell\footnote{This can be obtained from the disk
distribution by taking $\ell_1^2\rightarrow -\ell_1^2$ and $r\ge
\ell_1$, as in \cite{sfetsos}.} ($\ell_1=0$ and $\ell_2=\ell_3\ne
0$), the sigma deformations actually induce a phase transition
from a Coulombic phase for small distance to a linearly confining
phase as the distance becomes larger. This is shown in Figure 4
for the case of $\ell_1=0$, $\ell_2=\ell_3=10$ and $\hat\sigma=1$,
$2$ and $3$. For small distances between the quarks and
anti-quarks, they still experience Coulombic attraction, as in the
case of the conformal fixed point in the UV limit. As the distance
increases, this transitions to a regime of linear confinement,
where the scale of confinement increases with $\hat\sigma$. This
is very similar to what happens for certain string trajectories in
parts of the moduli space of the undeformed theory which exhibit a
confinement phase, such as case II in \cite{sfetsos1} as shown in
Figure 5 of that paper. However, in the limit of vanishing
$\sigma$, our trajectory matches that of case I in
\cite{sfetsos1}, for which there was no confining behavior.

In order to see the behavior for large length $L$, consider
(\ref{case1}) as $\alpha_2\rightarrow 1$, which implies that $r_0$
and $\ell_1$ are small. In this approximation, we get the simple
analytical result
%%%%
\be E = \fft{{\hat\sigma}\ell_2^2}{2\pi R^2}\,L\,. \ee
%%%%
This result can be obtained either by expanding the complete
elliptic integrals around $\alpha_2=1$, or equivalently by
assuming that the dominant contribution arises from the region
near $r=r_0$. As expected, for nonzero ${\hat\sigma}$ we have
linear confinement at large distance. Note that the scale of
confinement is independent of the coupling constant $g_{YM}$ in
this limit. In the cases considered in Figure 4, the typical
length scale at which the Coulombic behavior transitions to linear
confinement is on the order of $.2 R^2/\ell_2$.

For our deformed solution (\ref{background}), similar behavior can
also be found for a uniform distribution of D3-branes on a
particular family of five-dimensional ellipsoids (all $\ell_i$
nonvanishing and $\ell_2=\ell_3$). For the more general examples
with $SO(2)\times SO(2)\times SO(2)$ symmetry ($\ell_2\ne
\ell_3$), there are no longer any purely radial trajectories.
Nevertheless, we expect to find that these modes also undergo
transitions to confining phases. For nonradial trajectories,
$\gamma$ deformations also have an effect. Cases in which there is
linear as well as logarithmic confining behavior were studied in
\cite{sfetsos}.

The second case for which $E$ and $L$ can be written in terms of
complete elliptic integrals \cite{gr} is
%%%%
\bea L &=& \fft{R^2}{(\ell_1^2-\ell_2^2)}
\sqrt{\fft{r_0^2+\ell_2^2}{2\beta_1
(1+\hat\sigma^2)(r_0^2+\ell_1^2)}} \,
[\Pi (\ft{\ell_1^2-\ell_2^2}{r_0^2+\ell_1^2},\sqrt{q})-K(\sqrt{q})]\,,\nnn\\
E &=& \frac{1}{2\pi \beta_2 \sqrt{(1+
\hat\sigma^2)(r_0^2+\ell_1^2)}} \left[ \alpha_1 K(\sqrt{q})
-\frac{1}{(1-q)} E(\sqrt{q}) \right]\,,\label{case2} \eea
%%%%
where
%%%%
\be q \equiv \fft{\hat\sigma^2
\ell_1^2-(1+2\hat\sigma^2)\ell_2^2-(1+\hat\sigma^2)r_0^2}{(1+\hat\sigma^2)
(r_0^2+\ell_1^2)}\,. \ee
%%%%
This case applies when $r_0^2<\ell_1^2-2\ell_2^2$ and
$\hat\sigma^2> \ft{r_0^2+\ell_2^2}{\ell_1^2-2\ell_2^2-r_0^2}$.
Since $\hat\sigma$ is nonvanishing for this case, there is no
limit in which it reduces to undeformed results. We have found
similar behavior as shown in Figures 3 and 4 for the various
values of $\ell_i$ and $\sigma$.

\section{Discussion}

We have considered the gravity dual of deformations of the Coulomb
branch of ${\cal N}=4$ super Yang-Mills theory. In particular, for
$\sigma$ deformations we have used Wilson loops to probe the
geometry in cases which preserve $SO(4)\times SO(2)$ symmetry. In
certain parts of the moduli space, we have found that $\sigma$
deformations induce a transition from Coulombic attraction between
quarks and anti-quarks to linear confinement.

We have already noted that the supergravity description breaks
down for string configurations which get too close to the singular
shell of D3-branes. In this region, one can no longer approximate
the brane distribution as being continuous. This description must
be replaced by discretely-spaced stacks of D3-branes. We can now
consider a string getting close to a particular stack which has a
large number $n$ of D3-branes, where $1<<n<<N$. For the cases
considered in Figure 3, for which the Coulombic behavior persists
after the deformations, there is partial screening of charges
since $R^4$ is effectively reduced by a factor of $n/N$
\cite{sfetsos}. For cases in which there is linear confinement, as
presented in Figure 4, we find that the confinement is actually
occurring at an intermediate energy scale of the theory. In both
very high and low energy scales, there is Coulombic behavior.
There is partial charge screening in the deep IR region relative
to the UV limit.

The concavity condition on the potential of a heavy quark
anti-quark pair is given by
%%%%
\be \fft{dE}{dL}>0\,,\qquad \fft{d^2E}{dL^2}\le 0\,. \ee
%%%%
It can be seen directly from Figures 3 and 4 that these conditions
are obeyed. The first condition implies that the force between the
quark and anti-quark is always attractive. Although this is
obvious from QCD, this is not obvious from the viewpoint of the
string theory a priori. The second condition implies that the
force does not increase with $L$. The concavity condition is
sometimes not satisfied when the string configuration is too close
to the D3-brane shell, since then the continuous approximation of
the brane distribution breaks down \cite{sfetsos1}.

Although the $\beta$ deformations can have a dramatic effect on
the behavior of radial string configurations, the
minimally-coupled massless scalar field equation for purely radial
modes remains unchanged. This seems to indicate that the quarks
and glueballs in the deformed theory are associated with two
independent confinement scales\footnote{We are grateful to
Leopoldo Pando Zayas for bringing up this possibility}. On the
other hand, scalar modes with angular dependence are affected by
$\beta$ deformations. This was discussed for the case of $\gamma$
deformations in \cite{sfetsos}, where it was found that the wave
equation can be written as the Heun differential equation, which
is related to the BC$_1$ Inozemtsev system by way of a
non-standard trigonometric limit. Since this is an integrable
model, the Bethe ansatz method can be used to find solutions
\cite{sfetsos}. For the full $\beta$ deformations, the $H$ factors
from the background metric in (\ref{background}) cancel out of the
wave equation, regardless of angular dependence. From $G$ in
(\ref{G}), the wave equation has the same form as that for
$\sigma=0$ with the replacement $\gamma\rightarrow\beta$ and is
therefore integrable.

In the case of $\gamma$ deformations, it has been explicitly shown
that the deformed supergravity solution preserves ${\cal N}=1$
supersymmetry \cite{sfetsos2}. We expect that the same is true for
$\sigma$ deformations, since the direction corresponding to the
$U(1)$ R-symmetry is not involved in the deformation procedure.
However, it would be nice to show this from the Killing spinor
equations.

We have restricted ourselves to only purely radially oriented
string configurations. One could also consider strings whose
endpoints lie at different angles in the internal space. This
would correspond to giving different expectation values to the two
$U(1)$ factors in the global gauge group corresponding to the
quark and the anti-quark \cite{malda2}. Such configurations have
already been considered for the supergravity backgrounds
describing the undeformed Coulomb branch of ${\cal N}=4$ Yang
Mills in \cite{sfetsos1}, as well as for the $\gamma$ deformed
backgrounds in \cite{sfetsos}. However, one of the main points of
this paper is that $\sigma$ deformations, as opposed to $\gamma$
deformations, can have a radical effect on the behavior of purely
radial string configurations.

It would also be interesting to consider how these $\sigma$
deformations change the confining characteristics and phase
structure of finite temperature field theories described by
nonextremal rotating D3-branes.

Finally, in this paper we have only considered external quarks. It
would certainly be nice to consider dynamical quarks through the
addition of D7-branes.

\vspace{.7cm}

\centerline{\bf Acknowledgments}

We would like to thank Philip Argyres, Sangmin Lee, Hai Lin, Hong
L\"{u}, Oleg Lunin, Carlos N\'u\~nez, Chris Pope, Konstadinos
Sfetsos and Rohana Wijewardhana for useful discussions. We are
grateful to comments made by Konstadinos Sfetsos regarding the
earlier version of this paper. We thank the Institute for Advanced
Study for hospitality during the early stages of this work. C.A.
thanks the Korea Institute for Advanced Study (KIAS) where this
work was undertaken and his work was supported by grant No.
R01-2006-000-10965-0 from the Basic Research Program of the Korea
Science \& Engineering Foundation. The work of J.F.V.P. is
supported by DOE grant DOE-FG02-84ER-40153.

\end{document}